\documentclass[aps,prl,twocolumn,showpacs,groupedaddress, superscriptaddress, reprint, nofootinbib]{revtex4-1}
\usepackage{bbm,amsmath,amssymb,amsbsy,graphicx,subfigure,times}

\usepackage{hyperref,color}
\usepackage[latin1]{inputenc}
\newcommand{\be}{\begin{equation}}
\newcommand{\ee}{\end{equation}}
\newcommand{\ben}{\begin{eqnarray}}
\newcommand{\een}{\end{eqnarray}}
\newcommand{\bes}{\begin{subequations}}
\newcommand{\ees}{\end{subequations}}

\newcommand{\id}{\mathbbm{1}}

\newcommand{\tr}{{\rm Tr}\,}
\newcommand{\gr}[1]{\boldsymbol{#1}}

\newcommand{\eq}[1]{Eq.~(\ref{#1})}

\newcommand{\sig}{{\gr\sigma}}

\begin{document}
\title{Quantum versus classical correlations in Gaussian states}

\date{\today}
\author{Gerardo Adesso}
\affiliation{School of Mathematical Sciences, University of Nottingham,
University Park,  Nottingham NG7 2RD, UK.}

\author{Animesh Datta}
\affiliation{Clarendon Laboratory, Department of Physics, University of Oxford, OX1 3PU
Oxford, UK}

\begin{abstract}
Quantum discord, a measure of genuinely quantum correlations, is generalized to continuous variable systems. For all two-mode Gaussian states, we calculate analytically the quantum discord and a related measure of classical correlations, solving an optimization over all Gaussian measurements. Almost all two-mode Gaussian states are shown to have quantum correlations, while for separable states, the discord is smaller than unity. For a given amount of entanglement, it admits tight upper and lower bounds. Via a duality between entanglement and classical correlations, we derive a closed formula for the Gaussian entanglement of formation of all mixed three-mode Gaussian states whose normal mode decomposition includes two vacua.
\end{abstract}


\maketitle

Entanglement, non-classicality and non-locality are among the pivotal features of the quantum world.
While for pure quantum states these concepts are all equivalent, being like three facets of the same gemstone, they correspond to different resources in the general case of mixed states. Namely, while entanglement plays a central role in bipartite and multipartite quantum communication~\cite{hhh}, its necessity for mixed-state  quantum computation is still unclear~\cite{datta05a}. Conversely, several recent studies have shown that separable (that is, not entangled)  states, traditionally referred to as ``classically correlated'', might  retain  some signatures of quantumness with potential operational applications for quantum technology~\cite{dattabarbieri,piani,acin,terno}. One such signature is the {\it quantum discord}~\cite{zurek}, which strives at capturing all the quantum correlations in a bipartite state, including -- but not restricted to -- entanglement. Progresses in the demonstration of quantum protocols for the manipulation and transmission of information, as well as in the quantification of entanglement, which is closely related to the efficiency of such protocols,  have been amply recorded in the case of finite-dimensional systems, mainly qubits, as well as infinite-dimensional systems, where correlations between degrees of freedom with continuous spectra are exploited \cite{bl05}. However, there persists a  fundamental gap  between finite and infinite dimensional systems concerning the investigation of more general measures of quantumness versus classicality~\cite{ai07}. For Gaussian states, the workhorses of continuous variable quantum information, such an investigation would be especially valuable, since in view of the positivity of their Wigner distribution these states have been sometimes tagged as essentially classical.

In this Letter we endeavor to bridge this gap. We define the quantum discord for Gaussian states and explicitly solve the optimization problem involved in its definition, constrained to measurements that preserve the Gaussian character of the states. We derive a closed formula for the ensuing \emph{Gaussian quantum discord} and for a related measure of classical correlations~\cite{henderson01a} on {\it all} two-mode Gaussian states. We prove that these quantum correlations are limited for separable Gaussian states, yet they are nonzero for all but product states. For entangled states, quantum discord is found to admit tight upper and lower bounds, given by functions of the Gaussian entanglement of formation \cite{geof}. Exploiting a duality between entanglement and classical correlations~\cite{koashi}, we derive an analytical formula for the Gaussian entanglement of formation of all three-mode mixed Gaussian states that are reductions of four-mode pure Gaussian states. Our results unveil  the general structure and nature of correlations in continuous variable Gaussian states, and can find immediate applications in assessing the performance of quantum information primitives.

Quantum discord~\cite{zurek} originates from the discrepancy between two classically equivalent definitions of mutual information, a measure of total  correlations in a quantum state. For classical probability distributions, the quantities $I(A:B)=H(A)+H(B)-H(A,B)$, $J(A:B)=H(A)-H(A|B)$, and $J(B:A)=H(B)-H(B|A)$ all coincide due to Bayes' rule, where $H$ is the Shannon entropy and the conditional entropy $H(A|B)$ is an average of the Shannon entropies of $A$ conditioned on the alternatives of $B$.
For a bipartite quantum state $\varrho_{AB}$, the mutual information can be defined as $I(\varrho_{AB}) = S(\varrho_{A})+S(\varrho_{B})- S(\varrho_{AB})$, where $S$ stands for the Von Neumann entropy, $S(\varrho) = -\tr(\varrho \log \varrho)$ (throughout the paper, $\log$ denotes the natural logarithm). The quantum analogue of $J(A:B)$, known as {\it one-way classical correlation} and denoted as ${\cal J}^\leftarrow(\varrho_{AB})$, is operationally associated with the distillable common randomness between the two parties~\cite{devetak}, and
depends on the measurements $\{\Pi_i\}$, $\sum_i \Pi_i = \mathbbm{1}$, made on $B$~\cite{henderson01a}. The state of $A$ after the measurement is given by $\varrho_{A|i} = \tr_B(\varrho_{AB}\Pi_i)/p_i,\;p_i=\tr_{A,B}(\varrho_{AB}\Pi_i).$ A quantum analogue of the conditional entropy can then be defined as ${\cal H}_{\{\Pi_i\}}(A|B)\equiv\sum_ip_i S(\varrho_{A|i}),$ and the one-way classical correlation, maximized over all possible measurements, takes the form $\mathcal{J}^{\leftarrow}(\varrho_{AB}) =  S(\varrho_A)-\inf_{\{\Pi_i\}}{\cal H}_{\{\Pi_i\}}(A|B).$ The quantum discord is finally defined as total minus classical correlations:
\ben
\label{discexp}
\mathcal{D}^\leftarrow(\varrho_{AB}) &=& I(\varrho_{AB})-\mathcal{J}^\leftarrow(\varrho_{AB})\\  &=& S(\varrho_B)-S(\varrho_{AB})+\inf_{\{\Pi_i\}}{\cal H}_{\{\Pi_i\}}(A|B).\nonumber
\een
We denote by $\mathcal{J}^{\rightarrow}(\varrho_{AB})$ and $\mathcal{D}^\rightarrow(\varrho_{AB})$ the corresponding (generally different) quantities where the roles of $A$ and $B$ are swapped. On pure states, quantum discord coincides with the entropy of entanglement $S(\varrho_B)=S(\varrho_A)$. States with zero  discord represent essentially a classical probability distribution embedded in a quantum system, while  a positive  discord, even on separable (mixed) states, is an indicator of quantumness, that arises e.g.~when $\varrho_{AB}$ has entangled
eigenvectors~\cite{terno,dattainprep}, and may  operationally be associated to the impossibility of local broadcasting~\cite{piani}.

We now define a Gaussian version of quantum discord and calculate it analytically for all two-mode Gaussian states.
A two-mode Gaussian state $\varrho_{AB}$ is fully specified, up to local displacements, by its covariance matrix (CM) $\sig_{AB}$ of elements $\sigma_{ij} = {\rm Tr} [\varrho_{AB} \{\hat{R}_i,  \hat{R}_j\}_+]$ where ${\bf \hat{R}} = (\hat{x}_A,\hat{p}_A,\hat{x}_B,\hat{p}_B)$ is the vector of phase-space operators satisfying the canonical commutation relations $[\hat{R}_i, \hat{R}_j] = i \Omega_{ij}$, with $\gr \Omega$ being the symplectic matrix ${{\ 0\ 1}\choose{-1\ 0}}^{\oplus 2}$ \cite{ai07}. By means of local unitary (symplectic at the CM level) operations, every two-mode CM can be transformed in a standard form with diagonal subblocks
\begin{equation}\label{cm}
\sig_{AB}=\left(\begin{array}{cc}
\gr\alpha & \gr\gamma \\
\gr\gamma^T & \gr\beta
\end{array}\right)\quad \mbox{with} \quad
\begin{array}{l} \gr\alpha= a \id ,  \gr\beta=b \id , \\ \gr\gamma={\rm diag}\{c,d\}
\end{array}.
\end{equation}
Let us define the symplectic invariants $A=\det\gr\alpha$, $B=\det\gr\beta$, $C=\det\gr\gamma$, and $D=\det\sig_{AB}$.
The CM corresponds to a physical state iff  $A,B \ge 1$ and $\nu_{\pm} \ge 1,$ where the symplectic eigenvalues
are defined by $2\nu_{\pm}^2 = \Delta \pm \sqrt{\Delta^2-4D}$ with $\Delta = A+B+2C$. A Gaussian state with CM $\sig_{AB}$ is entangled iff $\tilde{\nu}_- <1$, where the smallest symplectic eigenvalue $\tilde{\nu}_-$ of the partially transposed CM is obtained from $\nu_-$ by replacing $C$ with $-C$, i.e. by time reversal \cite{simon}.

Both the one-way classical correlations ${\cal J}^\leftarrow$ and the quantum discord ${\cal D}^\leftarrow$ are entropic quantities, and therefore invariant under local unitaries. Hence, we can derive their closed formulae exploiting the standard form of a general two-mode Gaussian state, and later recast our results in terms of the four invariants of the state. The {\it Gaussian quantum discord} of a two-mode Gaussian state $\varrho_{AB}$ can be defined as the quantum discord where the conditional entropy is restricted to generalized Gaussain POVMs on $B$. These are all measurements executable using linear optics and homodyne detection \cite{giedkefiurasek}. We then have ${\cal D}^{\leftarrow}(\varrho_{AB}) = S(\varrho_B) - S(\varrho_{AB})+  \inf_{\Pi_B(\eta)} \int d{\eta} p_B(\eta) S(\varrho_{A_{\eta}})$.
Here the Gaussian measurement  $\Pi_B(\eta)$ on subsystem $B$ can be written in general as $\Pi_B(\eta) = \pi^{-1} \hat{W}_B(\eta) \Pi^0_B \hat{W}^\dagger_B(\eta)$ where $\hat{W}_B(\eta) = \exp(\eta \hat{b}^\dagger - \eta^\ast \hat{b})$ is the Weyl operator, $\hat{b} = (\hat{x}_B + i \hat{p}_B)/\sqrt2$, $\pi^{-1}\int d^2\eta \Pi_B(\eta) = {\mathbbm{1}}$ and $\Pi^0_B$ is the density matrix of a (generally mixed) single-mode Gaussian state. The conditional entropy is a concave function of the POVM elements, i.e. it is concave on the set of single-mode Gaussian states $\Pi^0_B$. Gaussian states do not form a convex set, yet  every Gaussian state admits  a convex decomposition into pure Gaussian states, so    it  sufficient (as in the finite dimensional case) to restrict to states $\Pi^0_B$ that are {\it pure}, single-mode Gaussian states \cite{alessio} whose CM we denote as $\sig_0$. The conditional state $\varrho_{A|\eta}$ of subsystem $A$ after the measurement $\Pi_B(\eta)$ on $B$ has  a CM independent of the measurement outcome \cite{giedkefiurasek}  and given by $\gr\varepsilon = \gr\alpha-\gr\gamma (\gr\beta+ \sig_0)^{-1} \gr\gamma^T$.
Recalling then that the Von Neumann entropy of a $n$-mode Gaussian state with CM $\sig$ can be computed as \cite{holevowerner} $S(\sig)=\sum_{i=1}^N f(\nu_i)$, where $\nu_i$ are the symplectic eigenvalues of the state and  $f(x) = \left(\frac{x+1}{2}\right) \log\left[\frac{x+1}{2}\right] -\left(\frac{x-1}{2}\right) \log\left[\frac{x-1}{2}\right]$,
the one-way classical correlation and the Gaussian quantum discord for two-mode Gaussian states with CM $\sig_{AB}$ are
\begin{eqnarray}
{\cal D}^{\leftarrow}(\sigma_{AB}) &=& f(\sqrt{B})-f(\nu_-) - f(\nu_+) + \inf_{\sig_0} f(\sqrt{\det{\gr\varepsilon}})\,. \nonumber \\
{\cal J}^{\leftarrow}(\sigma_{AB}) &=& f(\sqrt{A})-  \inf_{\sig_0} f(\sqrt{\det{\gr\varepsilon}})\,. \label{Gdiscord}
\end{eqnarray}
To get closed formulae we need to minimize $\det(\gr\varepsilon)$ over all CMs $\sig_0$ corresponding to pure one-mode Gaussian states, i.e. rotated squeezed states: $\sig_0 = R(\theta) {\rm diag}\{\lambda, 1/\lambda\} R^T(\theta)$, where $\lambda \ge 0$ and $R(\theta) = {{
\cos\theta \ \sin\theta}\choose{
 -\sin\theta\ \cos\theta}}$.
For a general two-mode Gaussian state in standard form, one has
$E(\lambda,\theta)\doteq\det(\gr{\varepsilon})= \big[2 a^2 (b + \lambda) (1 + b \lambda)-a \left(c^2+d^2\right) \left(2 b \lambda +\lambda ^2+1\right) +a \left(c^2-d^2\right) \left(\lambda ^2-1\right) \cos (2 \theta )+2 c^2 d^2 \lambda \big]/\big[{2 (b + \lambda) (1 + b \lambda)}\big]$.
We set without loss of generality $c \ge |d|$. We now look for stationary points of $E$ by studying its partial derivatives. One such point is  $\lambda= 1$, for which $\sig_0$ is the identity (regardless of $\theta$), and is  a saddle point except when $d=\pm c$.
Next, the equation $\partial_\lambda E = 0$ is quadratic in $\lambda$, but one of its roots is always negative. The other root, given by $\lambda=\lambda_2 =\big[a b  (d^2-c^2 )+c|d|\sqrt{ (a-a b^2+b c^2 ) (a-a b^2+b d^2 )}\big]/\big[a b^2 c^2- (a+b c^2 ) d^2\big]$ and $\theta=0$ (or, equivalently, $1/\lambda_2$ and $\theta=\pi/2$), is a local minimum and is acceptable  provided that $\lambda_2 \ge 0$, that is when  $\delta=\left(-a c_1^2+b \left(a b-c_1^2\right) c_2^2\right)  \ge 0$.
Additional candidates for $\inf E$ have to be sought at the boundaries of the parameter space: a potential minimum lies at $\lambda \rightarrow 0$, $\theta=0$ (or equivalently $\lambda \rightarrow \infty$, $\theta=\pi/2$).
In the whole physically allowed region for the CM parameters $a$, $b$, $c$ and $d$, we have  $E(1,0) \ge E(\lambda_2,0)$, with equality holding only when $d=\pm c$, and   $E(\lambda_2,0) \le E(0,0)$. Thus, for any ${\sig_{AB}}$,
$\inf_{\sig_0} \det(\gr\varepsilon)$ is equal to $E(\lambda_2,0)$ if $\delta\ge 0$ and to $E(0,0)$ otherwise.
In terms of symplectic invariants, the Gaussian quantum discord and the one-way classical correlation  for a general two-mode Gaussian state $\sig_{AB}$ are given  by Eq.~(\ref{Gdiscord}) with
\begin{eqnarray}\label{infdet}
& E^{\min}& =  \inf_{\sig_0} \det(\gr\varepsilon)= \\ &  &\hspace*{-.5cm} \left\{  \hspace*{-.5cm}  \begin{array}{rcl}
& &\begin{array}{c}\frac{{2 C^2+\left(-1+B\right) \left(-A+D\right)+2 |C| \sqrt{C^2+\left(-1+B\right) \left(-A+D\right)}}}{{\left(-1+B\right){}^2}}\end{array},\\ & &\qquad  \left (D-A B  \right) {}^2 \le \left (1 +   B \right) C^2 \left (A + D \right); \\ \\
& &\begin{array}{c}\frac{{A B-C^2+D-\sqrt{C^4+\left(-A B+D\right){}^2-2 C^2 \left(A B+D\right)}}}{{2 B}}\end{array}, \\ & & \qquad \hbox{otherwise,} \end{array} \right.
   \nonumber
\end{eqnarray}
Notice that this  only depends on $|C|$, i.e., entangled ($C<0$) and separable states are treated on equal footing.
For states falling in the second case of  \eq{infdet}, homodyne measurements (projections onto infinitely squeezed states, $\lambda=0$) on $B$ minimize the conditional entropy of $A$. An example is when
\begin{equation} \label{glemdualsym}
A=D=a^2,\,C=(1-B)/2,\,B=b^2\,,
\end{equation}
with $1 \le b \le 2a-1$, which is a mixed state of partial minimum uncertainty, i.e., one of its normal modes is the vacuum: $\nu_-=1$ \cite{ai07}.
On the other hand, the first case corresponds to a more general measurement, i.e., projection of mode $B$ onto a squeezed state with unbalanced, finite variances on $\hat{x}_B$ and $\hat{p}_B$. A notable class of states satisfying the first case are squeezed thermal states (including pure states), characterized by $d=\pm c$, for which the conditional entropy is in particular minimized by heterodyne measurements (projection onto  coherent states, $\lambda=1$).
In general, Gaussian quantum discord can be accessed experimentally by linear optics, and our finding provides the optimal measurements to verify quantum correlations given the CM of a two-mode Gaussian state.

We now analyze the relationships between classical correlations, quantum discord, separability and entanglement. For every entangled state the quantum discord is strictly positive (since $S(\varrho_B) - S(\varrho_{AB}) >0$). Almost all separable states in finite dimensions have also nonzero discord~\cite{acin}. In any dimension (including infinite dimensions under the constraint of finite mean energy), the states $\varrho_{AB}$ with zero discord are the ones that saturate the strong subadditivity inequality for the Von Neumann entropy on a tripartite state $\varrho_{ABC}$ where $C$ is an ancillary system realizing the measurements on $B$~\cite{dattainprep,hjpw04}. From the characterization of such states in the Gaussian scenario~\cite{petz} (see Supplementary Appendix A \cite{epaps} for more details), it follows that  the only two-mode Gaussian states with zero Gaussian quantum discord are product states $\sig_{AB} = \gr\alpha \oplus \gr\beta$, i.e., states with no correlations at all, that constitute a zero measure set. Quite remarkably, then, {\it all correlated two-mode Gaussian states have non-classical correlations} certified by a nonzero quantum discord.  This is in qualitative agreement with a recent study \cite{slater} where it has been demonstrated numerically, using the lack of a positive-definite Glauber-Sudarshan $P$ representation as a nonclassicality criterion, that essentially all two-mode Gaussian states are {\it a priori} nonclassical.

For Gaussian states with asymptotically diverging mean energy, however, interesting correlation structures arise. Consider the squeezed thermal state given by
\begin{equation}
\label{cmivette}
\begin{split}
&a=\cosh(2s),\,b=\cosh^2 r \cosh(2s) + \sinh^2 r,\,\\
&c=-d=\cosh r \sinh(2s)\,.
\end{split}
\end{equation}
For $r=0$, this is a pure two-mode squeezed vacuum state, whose entanglement is an increasing function of $s$. In the limit $r \rightarrow \infty$, it is asymptotically separable (but not in product form). Concerning the discord (minimized in this example by heterodyne detections), we find ${\cal D}^{\leftarrow}(\sig_{AB}) =f[\cosh^2 r \cosh(2s) + \sinh^2 r]-f[\cosh^2 r + \cosh(2s) \sinh^2 r] \xrightarrow{r,s\rightarrow \infty} 0$ and ${\cal D}^{\rightarrow}(\sig_{AB}) = f[\cosh(2s)]-f[\cosh^2 r + \cosh(2s) \sinh^2 r]+f[\cosh(2r)]\xrightarrow{r,s\rightarrow \infty} 1$. While these limiting values are associated with ideal, unnormalizable states, they can be approached arbitrarily close by physical Gaussian states with large, but finite mean energy. Hence,  surprisingly, there exist bipartite Gaussian states such that: (i) they are non-product states, with arbitrarily large correlation matrix $\gr\gamma$, yet have infinitesimal quantum discord; (ii) their quantum correlations can  be revealed by probing only one subsystem, but not the other.
Thus motivated, we have  explored the discord asymmetry for one million randomly generated (separable and entangled) two-mode Gaussian states. Let ${\cal D}^{\max}\!=\!\max\{{\cal D}^{\leftarrow},{\cal D}^{\rightarrow}\}$, ${\cal D}^{\min}\!=\!\min\{{\cal D}^{\leftarrow},{\cal D}^{\rightarrow}\}$ for a given CM. We find  numerically that ${\cal D}^{\max}-{\cal D}^{\min}  \le {\cal D}^{\min}/[\exp( {\cal D}^{\min})-1] \le  1$. The leftmost bound is saturated by states of \eq{cmivette} in the limit $s \rightarrow \infty$, and unity is reached for $r \rightarrow \infty$ as well. The maximum discord asymmetry decays exponentially with  ${\cal D}^{\min}$, so when the discord calculated in either way is  large, we have {\it de facto} ${\cal D}^{\leftarrow}={\cal D}^{\rightarrow}$.

Next we ask: To what extent can separable Gaussian states be quantumly correlated? While their discord is typically nonzero (but for product states), we find that it cannot exceed {\it one unit} of information [Fig.~\ref{figdisc}(left)]. In the Supplementary Appendix B~\cite{epaps} we prove that for all two-mode separable Gaussian states, ${\cal D}^{\leftarrow}(\sig_{AB}^{sep}) \le [(b-1)/2] \log[(b+1)/(b-1)] \le 1$. The first inequality is saturated by separable squeezed thermal states whose correlation matrix has maximum determinant $C$ and whose CM has maximum asymmetry between the two modes: $c=d=1+ab-a-b, a\rightarrow \infty$ [solid (red) curve in Fig.~\ref{figdisc}(left)]. The second bound is reached for $b \rightarrow \infty$. This result immediately implies a sufficient condition for the entanglement of Gaussian states given their discord: {\it if ${\cal D}^{\leftarrow}(\sig_{AB}) > 1,$ then $\sig_{AB}$ is entangled}.

We now focus  on entangled states, and study how Gaussian quantum discord compares {\it quantitatively} to the entanglement of the state, specifically measured by the most ``compatible'' measure available, the Gaussian entanglement of formation (Gaussian EoF) ${\cal E}_G$ \cite{geof} This is defined for Gaussian states $\varrho_{AB}$ as the convex roof of the Von Neumann entropy of entanglement, restricted to decompositions of $\varrho_{AB}$ into pure Gaussian states. It can be evaluated via a minimization over CMs:  ${\cal E}_G(\sig_{AB}) = \inf_{\sig'_{AB} \le \sig_{AB}\ :\ {\det(\sig'_{AB})=1}} f(\sqrt{\det{\gr\alpha'}})$, where the infimum runs over all pure bipartite Gaussian states with CM  $\sig'_{AB}$ smaller than $\sig_{AB}$, and $\gr\alpha'$ is the reduction of $\sig'_{AB}$ corresponding to the marginal state of mode $A$. Compact formulae for ${\cal E}_G$ exist for all symmetric two-mode states  (where the Gaussian EoF coincides with the true EoF as the Gaussian decomposition is optimal) \cite{eofsym}, as well as asymmetric ones~\cite{geof,ordering}. In Fig.~\ref{figdisc}(right), we plot $\cal{D}^{\leftarrow}$ vs ${\cal E}_G$ for 30000 randomly generated  two-mode Gaussian states. We find that for a given  entanglement degree, the  discord is bounded {\it both from above and below}. To find the upper bound analytically, we can restrict, as in the separable case, to squeezed thermal states (see \cite{epaps}) with $d=-c$. Further optimization within this family of states
yields that, for all two-mode entangled Gaussian states, the quantum discord satisfies:
\begin{equation}\label{upper}
{\cal D}^{\leftarrow}(\sig_{AB})  \le \max\{{\cal E}_G(\sig_{AB}), 2 \cosh^2 r \log(\coth r)\}\,,
\end{equation}
where ${\cal E}_G(\sig_{AB}) = f(1+2 \sinh^{-2} r)$ implicitly defines $r$. The rightmost bound [solid (red) curve in Fig.~\ref{figdisc}(right)] dominates in the low entanglement regime (${\cal E}_G < 2 \log 2$) and corresponds to ${\cal D}^{\rightarrow}$ of the states of \eq{cmivette} in the limit $s \rightarrow \infty$. The leftmost bound in \eq{upper} [dotted (green) line in Fig.~\ref{figdisc}(right)] is instead reached on pure states, and sets an upper limit to the quantum discord of all two-mode Gaussian states with sufficiently high entanglement. On the other hand, for a given ${\cal E}_G$, the Gaussian discord also satisfies ${\cal D}^{\leftarrow}(\sig_{AB})  \ge  2  \log(\coth r)$, with $r$ as before. This lower bound follows from the fact that states of \eq{cmivette} with $s \rightarrow \infty$ are extremal for the discord asymmetry, and corresponds to ${\cal D}^{\leftarrow}$ of those states [dashed (blue) line in Fig.~\ref{figdisc}(right)]. Interestingly this entails that, asymptotically, {\it for all two-mode Gaussian states with ${\cal E}_G \gg 0$, their Gaussian quantum discord lies between ${\cal E}_G-1$ and ${\cal E}_G$}.

\begin{figure}[t]
\includegraphics[width=8.5cm]{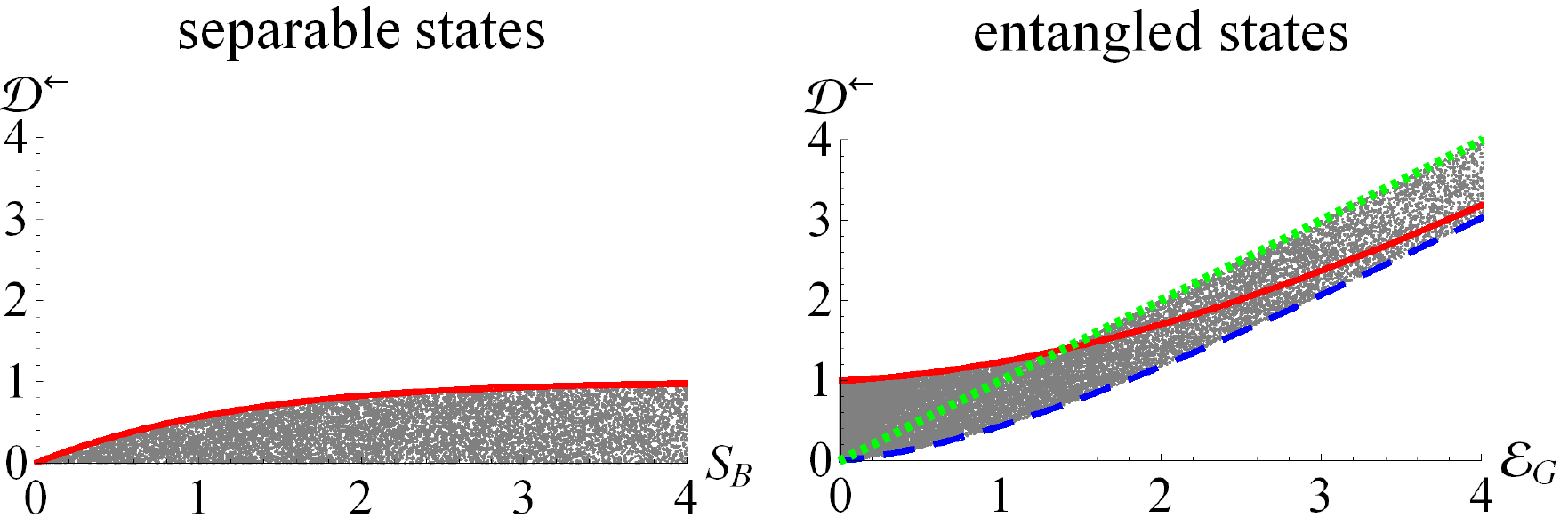}
\caption{(Color online). Left: Gaussian quantum discord versus marginal entropy for separable two-mode Gaussian states. Right: Gaussian quantum discord versus Gaussian EoF for entangled two-mode Gaussian states. See text for details of the bounding curves.}\label{figdisc}
\end{figure}

A further key result of our study is that \eq{infdet} provides a closed, computable formula for the Gaussian EoF of a class of {\it three-mode mixed Gaussian states}.
This is possible thanks to a duality relation between (Gaussian) classical correlations and (Gaussian) EoF~\cite{koashi} (see also~\cite{jenswolf}). Let $\varrho_{ABST}$ be a purification of $\varrho_{AB}$, i.e. a pure (Gaussian) state such that ${\rm Tr}_{ST}[\varrho_{ABST}] = \varrho_{AB}$  (we need in general two ancillary modes $S$ and $T$ to construct such a purification \cite{holevowerner}). Then: ${\cal J}^{\leftarrow}(\varrho_{AB}) + {\cal E}(\varrho_{AST}) = S(\varrho_A)$ where ${\cal E}(\varrho_{AST})$ denotes the EoF between party $A$ and the block of modes $ST$. In the Gaussian framework, from \eq{Gdiscord} we have simply: ${\cal E}_G(\sig_{AST}) = \inf_{\sig_0} f(\sqrt{\det{\gr\varepsilon}})$.
The states with CM $\sig_{AST}$ encompass all three-mode Gaussian states that are reductions of a four-mode pure Gaussian state. Their symplectic spectrum is of the form $\{1,1,b\}$, i.e. they are mixed states of partial minimum uncertainty, with two vacua as normal modes. For {\it all} such states, we now present an analytic method to compute the Gaussian EoF across the bipartition $A \times (ST)$: first, construct a purification, i.e. append an ancillary mode $B$ (with $\det(\sig_B)=\det(\sig_{AST})=b^2$) such that $\sig_{ABST}$ is pure \cite{holevowerner,geof}. Then, evaluate $E^{\min}$ of the marginal state $\sig_{AB}$ from \eq{infdet}. Finally, $f(\sqrt{E^{\min}})$ is the Gaussian EoF between $A$ and $ST$. An example of such a state, of relevance in a cryptographic setting, is provided in \cite{epaps} (Appendix C).

Finally, we wish to point out that for states $\sig_{AB}$ with $\nu_-=1$, the purification requires a single ancillary mode $S$, and the Gaussian EoF between modes $A$ and $S$, as computed through \eq{infdet}, agrees with the formula derived in \cite{ordering}. In the special case of these states given by \eq{glemdualsym}, the complementary state of modes $AS$ is symmetric. Thus the optimality of Gaussian decompositions for the EoF \cite{eofsym} implies that no non-Gaussian measurement can do better than the optimal Gaussian POVM devised here for the optimization of the conditional entropy. In other words {\it the (unrestricted) quantum discord and one-way  classical correlations in the  family of two-mode states $\sig_{AB}$ of \eq{glemdualsym} coincide with their Gaussian counterparts}. For general two-mode Gaussian states, it is an open question whether non-Gaussian measurements (e.g.~photo-detection) can lead to a further minimization of the  discord. Insights on this issue might be drawn if the additivity conjecture for bosonic channels were proven \cite{jenswolf}.

This Letter opens the way for the study of quantum discord in general multimode correlated bosonic systems,  and through the paradigmatic two-mode case demonstrates the `truly quantum' nature of Gaussian states, reinforcing their key role in continuous variable quantum information processing. Such studies could also lead to a deeper understanding of entanglement and the general structure of distributed correlations in harmonic lattices.  \quad

{\noindent {\it Acknowledgements.---} We thank M. Guta  for discussions. AD was supported by the EPSRC grant EP/C546237/1 and the EU Integrated Project QAP, and was at Imperial College, London when this work was completed.

{\noindent \it Note added.---} After the completion of this study, during the writing of the present paper, another work appeared \cite{giorda} where Gaussian quantum discord is independently defined for Gaussian states and explicitly calculated only in the specific case of two-mode squeezed thermal states.


\clearpage

\appendix
\section{Supplementary Material}

\begin{center}
{\bf Quantum versus classical correlations in Gaussian states}
 \quad \\ \quad \\

Gerardo Adesso and Animesh Datta
\end{center}

\subsection{A. Gaussian states saturating the strong subadditivity of entropy}

For a general state $\varrho_{ABC}$ of a tripartite system, the Von Neumann entropy is strongly subadditive:
$$S(\varrho_{AB}) + S(\varrho_{BC}) \ge S(\varrho_{ABC}) + S(\varrho_B) \,.$$
For finite-dimensional and infinite-dimensional systems (in the latter case, under the restriction of finite mean energy), it can be proven that  the states $\varrho_{AB}$ with zero quantum discord ${\cal D}$ are obtained from tripartite states  $\varrho_{ABC} \neq \varrho_{AB} \otimes \varrho_C$ that saturate the strong subadditivity. Here  $C$ is an ancillary system needed to realize the most general measurement on subsystem $B$, and $\tr_C[\varrho_{ABC}]=\varrho_{AB}.$ \cite{hjpw04,dattainprep}. In the Gaussian scenario, $C$ is in general a collection of Gaussian modes, and interactions are restricted to Gaussian ones, i.e., symplectic operations and homodyne detection, that preserve the Gaussian character of the involved states. Therefore, the Gaussian states with CM $\sig_{AB}$ that have zero Gaussian quantum discord are the ones such that strong subadditivity is saturated on Gaussian extensions $\sig_{ABC} \neq \sig_{AB} \oplus \sig_C$.
In the general case of tripartite Gaussian states where each subsystem $A$, $B$, and $C$ contains an arbitrary number of modes, Petz and Pitrik have recently characterized the subset of Gaussian states (tagged as Markov states) saturating the strong subadditivity  \cite{petz}. For a tripartite $n$-mode CM
$$\sig_{ABC} \equiv \sig_{\underbrace{\alpha_1,\ldots,\alpha_l}_A, \underbrace{\alpha_{l+1},\ldots\alpha_m}_B, \underbrace{\alpha_{m+1},\ldots\alpha_n}_C}
$$
corresponds to a Gaussian Markov state if and only if there exists a splitting within the block of modes $B$ (including the boundaries) such that the CM is block-diagonal, i.e., if and only if $$\sig_{ABC} = \sig_{\alpha_1, \ldots, \alpha_{l+j}} \oplus \sig_{\alpha_{l+j+1}, \ldots, \alpha_n}$$ for some $j$, $0 \le j \le m-l$.
In our setting, the block $B$ is made of a single mode only, hence the only possibilities are $j=0,1$. Consequently, the only Gaussian Markov states in this case are either of the form $\sig_{ABC} = \sig_{A} \oplus \sig_{BC}$ or $\sig_{ABC} = \sig_{AB} \oplus \sig_C$. The latter  having been discounted already, it follows that the only zero-discord two-mode Gaussian states (under the assumption of finite mean energy) are the reductions of $\sig_{A} \oplus \sig_{BC}$ after partial trace on $C$, i.e., product states $\sig_{AB} = \sig_A \oplus \sig_B$.

\subsection{B. Upper bound on the quantum discord of separable Gaussian states at given marginal entropy}
Here we prove the upper bound on quantum discord for separable Gaussian states at fixed $B=b^2$. First, we observe that, since the discord ${\cal D}^\leftarrow(\sig_{AB}) \equiv {\cal D}(A,B,C,D)$ (notice the slimmed notation and the explicit dependence on the invariants) involves a minimization over the Gaussian POVMs on system $B$, any such measurement provides an upper bound for the discord. In particular, heterodyne measurements do the job, being ${\cal D}(A,B,C,D) \le {\cal D}_{hetero}(A,B,C,D)$ for all two-mode Gaussian states. Since we know that squeezed thermal states are the only ones that saturate the bound, we can restrict to them for the maximization of the discord, and write that for fixed $A$, $B$, and $C$, it is ${\cal D}(A,B,C,D) \le {\cal D}(A,B,C,(\sqrt{AB} \pm C)^2)$. The $\pm$ sign distinguishes between the two cases, $d=\mp c$, that can be associated to separable squeezed thermal states. Since the conditional entropy, from \eq{infdet}, does not depend on the sign of $C$, the maximal discord is obtained in the case of minimal global entropy $S(\sig_{AB})$. From the expression of the symplectic eigenvalues $\nu_\pm$ and the form of the concave function $f(x)$ we find that the case $d=c$, i.e. $D=(\sqrt{AB}-C)^2$ maximizes the global entropy. Next, we observe that the discord for the family of separable squeezed thermal states under investigation is a monotonically increasing function of $C$, hence it is maximized at the boundary $C=1-\sqrt{A}-\sqrt{B}+\sqrt{AB}$. We have thus proven that, for all two-mode separable Gaussian states, ${\cal D}(A,B,C,D) \le {\cal D}(a^2,b^2,1-a-b+ab,(a+b-1)^2) = f(b) - f(a+b-1)+f[(2a+b-1)/(1+b)]$. This is by itself an useful upper bound that depends on both local entropies of the two-mode state. To get a function of $b$ only, we observe that the obtained bound is a monotonically increasing function of $a$, and is hence maximized at $a \rightarrow \infty$. Using the fact that $f(y)-f(x) \xrightarrow{x,y \rightarrow \infty} \log(y/x)$, we finally find the bound in the text:
${\cal D}^\leftarrow(\sig_{AB}^{sep}) \le [(b-1)/2] \log[(b+1)/(b-1)]$ which is an increasing function of $b$ converging to $1$ for $b \rightarrow \infty$.

\subsection{C. Example of a three-mode mixed Gaussian state whose Gaussian EoF can be computed exactly}

We now provide an example of a three-mode family of states with CM $\sig_{AST}$, of relevance in a cryptographic setting, whose normal mode decomposition accommodates two vacua. Suppose that Sally and Tom are sharing a pure two-mode squeezed state of modes $S$ and $T$, $\sig^{in}_{ST}$, of the form  \eq{cm} with $a=b\equiv s, c=-d \equiv\sqrt{s^2-1}$, as a quantum communication channel. A malicious Eve remotely prepares a two-mode squeezed state $\sig^{in}_{AB}$ of her modes $A$ and $B$, given by $a=b\equiv n, c=-d\equiv\sqrt{n^2-1}$. Eve then, maintaining local access to her mode $B$, attempts to intercept the communication by sending her mode $A$ to interfere with Sally's mode $S$ through a beam-splitter with transmittivity $t$, as schematically depicted in Fig.~\ref{figeve}. After the attack, the global state of the four involved modes has a CM given by $\sig_{ABST} = {\cal B}_{AS}(t) (\sig^{in}_{AB} \oplus \sig^{in}_{ST}) {\cal B}_{AS}(t)$, where ${\cal B}_{ij}(t)$ can be expressed in the standard form of \eq{cm} with $a=-b\equiv \sqrt t, c=d=\sqrt{1-t}$. Now Eve's ability to eavesdrop depends crucially on the entanglement between her probe mode $A$ and the target modes $ST$.
The reduced state $\sig_{AST}$ has symplectic spectrum $\{1,1,n\}$ so it belongs to the class whose Gaussian EoF can be calculated thanks to our result. We find that ${\cal E}_G(\sig_{AST}) = f[t+a(1-t)]$. This is equal to the conditional entropy of the probe mode $A$ after Eve performs a heterodyne measurement (optimal in this case) on her local mode $B$. In other words, Eve generates quantum correlations with the communicating parties, no matter how weak the entanglement (quantified by $n$) she had established between her modes $AB$ in the first place. For $t \rightarrow 0$, there is an entanglement swapping between Eve's modes and Sally and Tom's modes, yielding the most destructive attack.
In the limit of two perfect quantum channels, $a,n \rightarrow \infty$, Eve is sacrificing the (infinite) entanglement between her two modes, spending it to get (infinite) entanglement with the two spied parties: yet, some quantum correlations persist between $A$ and $B$, being ${\cal D}^{\leftarrow}(\sig_{AB}) = 1 - \log 2$ in this limit.

\begin{figure}[t]
\includegraphics[width=5.5cm]{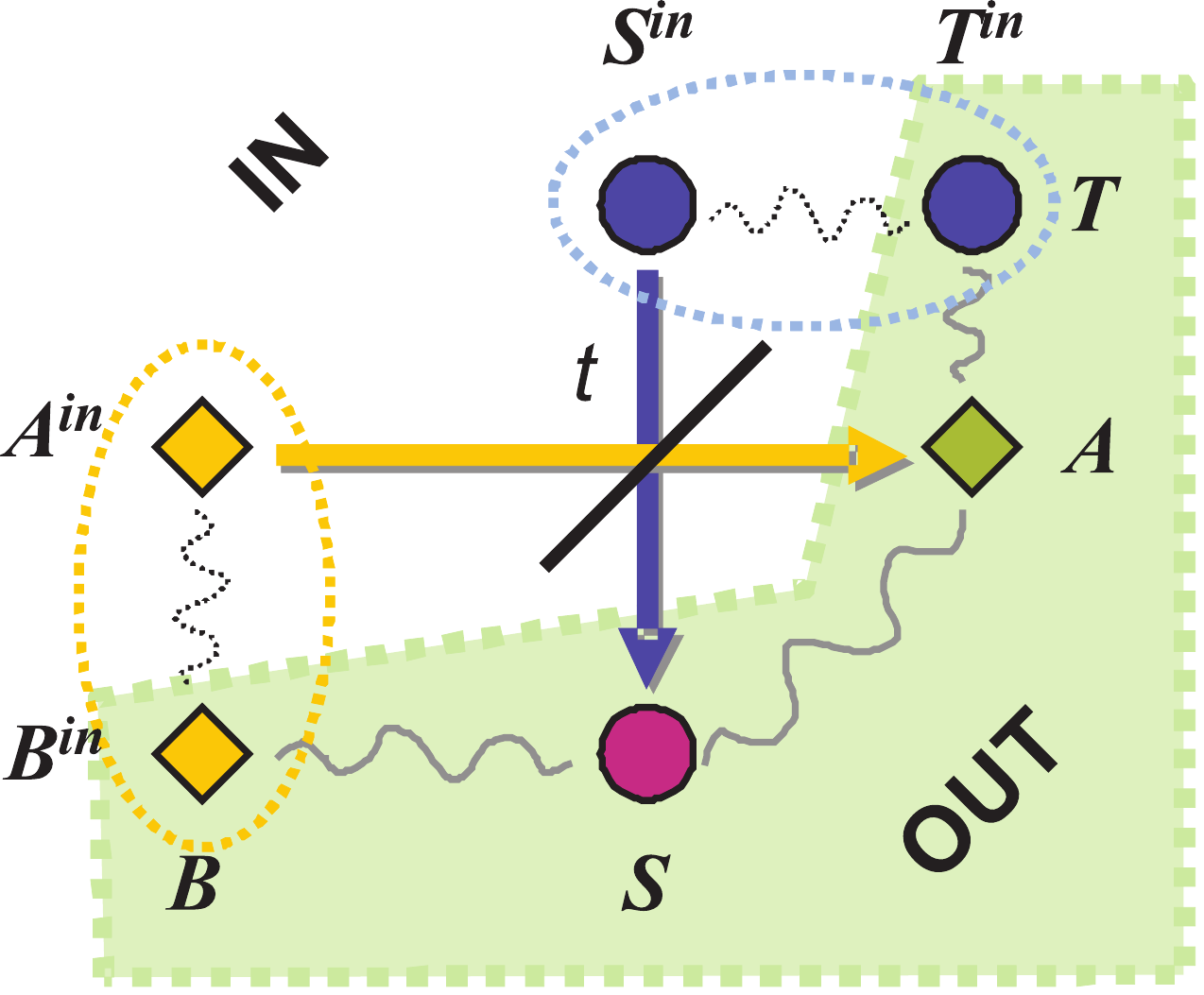}
\caption{Schematic picture of the production of the four-mode $\sig_{ABST}$ discussed in the text. Sally and Tom control modes $S$ and $T$, respectively, while Eve owns modes $A$ and $B$.}\label{figeve}
\end{figure}

\end{document}